\documentclass[prl,showpacs,twocolumn,superscriptaddress]{revtex4}
\usepackage{graphicx,latexsym,amssymb}
\usepackage{epstopdf}
\usepackage{amsmath}
\usepackage{amsfonts}
\usepackage{times}
\usepackage[colorlinks={true}]{hyperref}
\hypersetup{citecolor={blue}, filecolor={blue}, linkcolor={blue}, urlcolor={blue}}

\begin{document}

\title{Time-invariant entanglement and sudden death of non-locality}

\author{Bi-Heng Liu}
\affiliation{Key Laboratory of Quantum Information, University of Science and Technology of China, CAS,
Hefei, 230026, People's Republic of China}
\affiliation{Synergetic Innovation Center of Quantum Information and Quantum Physics, University of Science
and Technology of China, Hefei, 230026, People's Republic of China}

\author{Xiao-Min Hu}
\affiliation{Key Laboratory of Quantum Information, University of Science and Technology of China, CAS,
Hefei, 230026, People's Republic of China}
\affiliation{Synergetic Innovation Center of Quantum Information and Quantum Physics, University of Science
and Technology of China, Hefei, 230026, People's Republic of China}

\author{Jiang-Shan Chen}
\affiliation{Key Laboratory of Quantum Information, University of Science and Technology of China, CAS,
Hefei, 230026, People's Republic of China}
\affiliation{Synergetic Innovation Center of Quantum Information and Quantum Physics, University of Science
and Technology of China, Hefei, 230026, People's Republic of China}

\author{Chao Zhang}
\affiliation{Key Laboratory of Quantum Information, University of Science and Technology of China, CAS,
Hefei, 230026, People's Republic of China}
\affiliation{Synergetic Innovation Center of Quantum Information and Quantum Physics, University of Science
and Technology of China, Hefei, 230026, People's Republic of China}

\author{Yun-Feng Huang}
\affiliation{Key Laboratory of Quantum Information, University of Science and Technology of China, CAS,
Hefei, 230026, People's Republic of China}
\affiliation{Synergetic Innovation Center of Quantum Information and Quantum Physics, University of Science
and Technology of China, Hefei, 230026, People's Republic of China}

\author{Chuan-Feng Li}
\email{cfli@ustc.edu.cn}
\affiliation{Key Laboratory of Quantum Information, University of Science and Technology of China, CAS,
Hefei, 230026, People's Republic of China}
\affiliation{Synergetic Innovation Center of Quantum Information and Quantum Physics, University of Science
and Technology of China, Hefei, 230026, People's Republic of China}

\author{Guang-Can Guo}
\affiliation{Key Laboratory of Quantum Information, University of Science and Technology of China, CAS,
Hefei, 230026, People's Republic of China}
\affiliation{Synergetic Innovation Center of Quantum Information and Quantum Physics, University of Science
and Technology of China, Hefei, 230026, People's Republic of China}

\author{G\"{o}ktu\u{g} Karpat}
\affiliation{Turku Center for Quantum Physics, Department of Physics and Astronomy, University of Turku, FIN-20014 Turku, Finland}
\affiliation{Faculdade de Ci\^encias, UNESP - Universidade Estadual Paulista, Bauru, SP, 17033-360, Brazil}

\author{Felipe F. Fanchini}
\affiliation{Faculdade de Ci\^encias, UNESP - Universidade Estadual Paulista, Bauru, SP, 17033-360, Brazil}

\author{Jyrki Piilo}
\affiliation{Turku Center for Quantum Physics, Department of Physics and Astronomy, University of Turku, FIN-20014 Turku, Finland}

\author{Sabrina Maniscalco}
\email{smanis@utu.fi}
\affiliation{Turku Center for Quantum Physics, Department of Physics and Astronomy, University of Turku, FIN-20014 Turku, Finland}


\begin{abstract}
We investigate both theoretically and experimentally the dynamics of entanglement and non-locality for two qubits immersed in a global pure dephasing environment.  We demonstrate the existence of a class of states for which entanglement is forever frozen during the dynamics, even if the state of the system does evolve. At the same time non-local correlations, quantified by the violation of the Clauser-Horne-Shimony-Holt (CHSH) inequality, either undergo sudden death or are trapped during the dynamics.
\end{abstract}

\maketitle

Understanding correlations of quantum nature is essential both for fundamentals of quantum theory and for applications of quantum information science. Recent studies revealed that there exists different types of non-classical correlations in quantum systems, and they all prove to be relevant in the implementation of various tasks \cite{entrev,discrev}. Among them, entanglement can be argued to be the most fundamental, which is the resource of quantum computation and quantum information \cite{entrev}. Realistic quantum systems are, however, sensitive to their surroundings \cite{openbook} and, as a consequence, their characteristic traits tend to rapidly disappear.

Given that a quantum computer has to be resilient against the destructive effects of the environment, strategies aimed at preserving entanglement as long as possible are of great practical importance. Motivated by this challenge, several methods have been proposed to protect the correlations in the system from the noise. Examples are memory effects stemming from non-Markovian environments \cite{entnonmark}, dynamical decouplings techniques \cite{entdd} and quantum Zeno effect \cite{entzeno}. In addition, it has been shown that certain quantum correlations, e.g. quantum discord, might forever freeze throughout the dynamics under a suitable noise setting, becoming time-invariant \cite{frozendisc}. On the other hand, even though the phenomenon of time-invariant entanglement has been first noticed for a qubit-qutrit system for global dephasing noise \cite{frozenent1}, a more complete characterization has been obtained recently under a general global dephasing scenario \cite{frozenent2}.

Another manifestation of non-classical correlations is related to the quantum non-locality, stating that the predictions of quantum theory cannot be simulated by a local hidden variable model. For bipartite systems, non-local correlations can be identified through the violation of Bell inequalities \cite{bell}. All pure entangled states of two qubits violate a Bell inequality \cite{gisin} but this no longer holds for mixed states \cite{werner}. In particular, although entanglement is necessary for the existence of non-local correlations, there are entangled mixed bipartite states that do not violate Bell inequalities. To investigate whether quantum correlations quantified by entanglement are non-local or not, one can use a simple version of the Bell inequalities in the form of a Clauser-Horne-Shimony-Holt inequality \cite{chsh}. The dynamics of CHSH violation in relation with entanglement under decoherence has been studied theoretically \cite{kofman, entnonloc, entchsh,entchsh2,entchsh3} and experimentally \cite{Shu2010}. However, despite the studies in the literature on the relation of entanglement and non-locality, our results are distinct in the sense that they shed light on a fundamental difference of these two precious resources in a dynamical context. Also, if one considers quantum non-locality as the source of secure communication \cite{key}, our findings may stimulate new research on how to preserve non-local correlations in open systems.

Here we first theoretically study the dynamics of entanglement and quantum non-locality, as quantified by concurrence and CHSH inequality violation, respectively, for Bell-diagonal states under global dephasing. We demonstrate that while entanglement remains time-invariant for a particular class of Bell-diagonal states, the degree of violation of quantum non-locality can suffer sudden death and disappear in a finite time. Then, we report on an experiment with photonic qubits demonstrating our theoretical results. 

We consider two qubits globally interacting with stochastic dephasing noise along the $z$-direction, with the Hamiltonian $H(t)= -\frac{1}{2} n(t)(\sigma_{z}^A \otimes I^B + I^A \otimes \sigma_{z}^B)$, where $\sigma_{z}$ is the Pauli operator, $I$ is the identity operator, $n(t)$ is a stochastic field satisfying $\langle n(t) \rangle = 0$ and $\langle n(t)n(t') \rangle = \Gamma \delta(t-t')$, and $\Gamma$ is the damping rate associated with the field $n(t)$. The resulting dynamics of the system reads \cite{yueb}
\begin{eqnarray} \label{dyn}
\rho(t)=
\left(\begin{array}{cccc}
\rho_{11} & \rho_{12}\gamma & \rho_{13}\gamma & \rho_{14}\gamma^{4} \\
\rho_{21}\gamma & \rho_{22} & \rho_{23} & \rho_{24}\gamma \\
\rho_{31}\gamma & \rho_{32} & \rho_{33} & \rho_{34}\gamma \\
\rho_{41}\gamma^{4} & \rho_{42}\gamma & \rho_{43}\gamma & \rho_{44} \\
\end{array}\right),
\end{eqnarray}
where $\rho_{ij}$ denotes elements of the initial density matrix and the decoherence factor is $\gamma(t)= e^{ -t \Gamma /2}$.

\begin{figure}[t]
\includegraphics[width=0.28\textwidth]{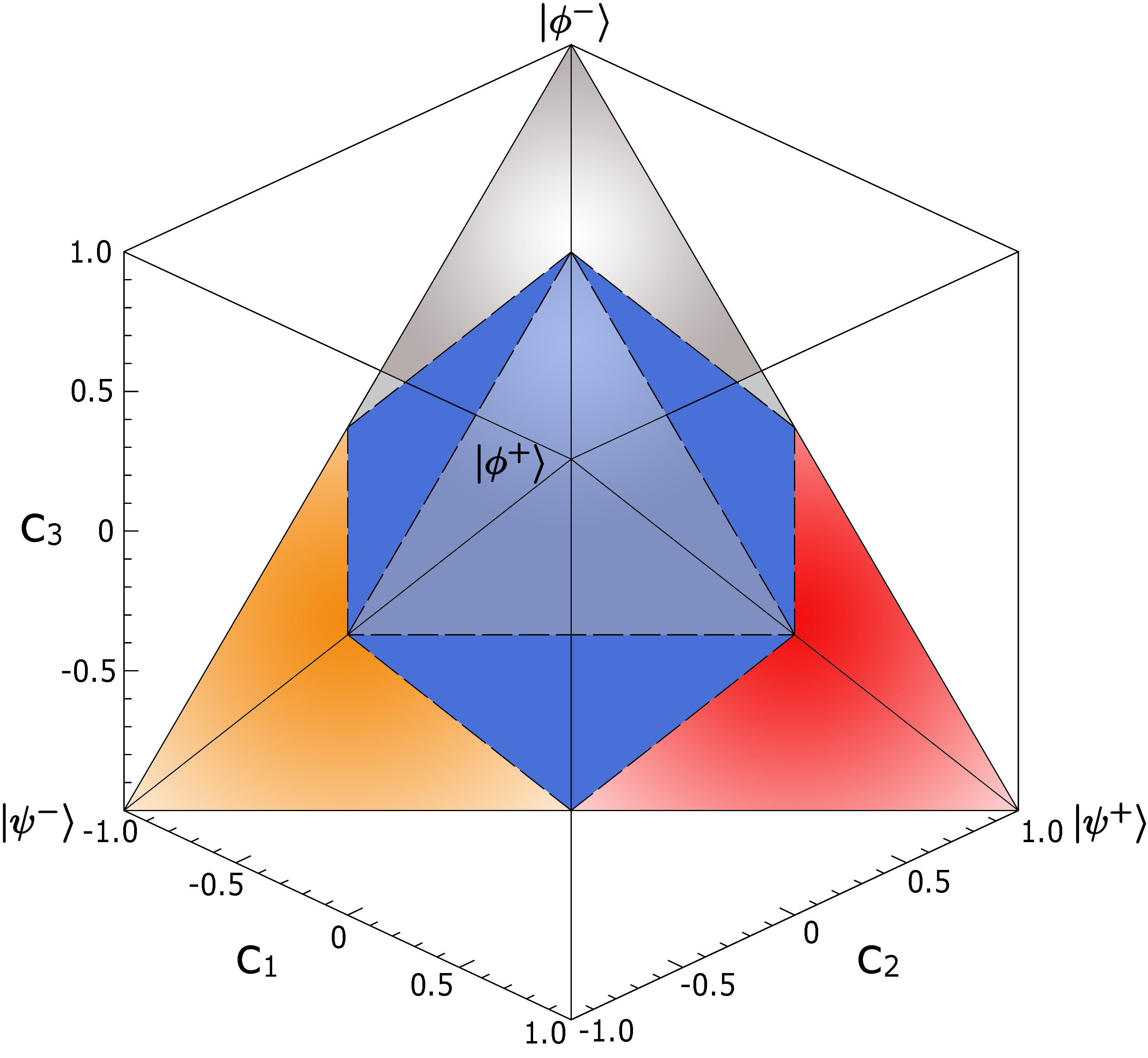}
\caption{The blue octahedron in the middle, marked with dashed lines, represents separable Bell-diagonal states. The states filling the red and orange regions,
connected to the Bell states $|\psi^+\rangle$ and $|\psi^-\rangle$ respectively, give rise to time-invariant entanglement.}
\label{fig1}
\end{figure}

We now introduce the family of two qubits states that have maximally mixed reduced density matrices.
Such states are known as Bell diagonal states and given as $\rho = (1/4) (I\otimes I +
\sum_{j=1}^3 c_j \sigma_j \otimes \sigma_j )$, where $\sigma_j$ are the Pauli operators, $c_j$ are real
numbers such that $0\leq |c_j| \leq1$, and the eigenvalues $(\lambda_k\geq0)$ of $\rho$ are given by $\lambda_{1,2} = \frac{1}{4}(1\mp c_1\mp c_2-c_3)$, $\lambda_{3,4} = \frac{1}{4}(1\pm c_1 \mp c_2+c_3)$. Bell diagonal states can be visualized as forming a tetrahedron with the four Bell states $|\phi^\pm\rangle=(|HH\rangle\pm| VV\rangle)/\sqrt{2}$ and $|\psi^\pm\rangle=(|HV\rangle\pm| VH\rangle)/\sqrt{2}$ sitting in the extreme points, as depicted in Fig. \ref{fig1}. We observe that our global dephasing map preserves the form of the initial states and transforms their coefficients as follows:
\begin{align} \label{cof}
c_{1}(t) &= \frac{1}{2} [c_{1}(1+\gamma^4(t))+c_2(1-\gamma^4(t))], \nonumber \\
c_{2}(t) &= \frac{1}{2} [c_{1}(1-\gamma^4(t))+c_2(1+\gamma^4(t))], \nonumber \\
c_{3}(t) &= c_{3} .
\end{align}
Consequently, $c_3(t)$ and $[c_1(t)+c_2(t)]$ remain invariant throughout the dynamics, while both
$c_1(t)$ and $c_2(t)$ asymptotically evolve to the same value, that is, $(c_1+c_2)/2$.

Concurrence of the Bell diagonal states are given as \cite{conc}
\begin{align} \label{ent}
E(\rho)=1/2 \max\{0,|c_1|+|c_2|+|c_3|-1\},
\end{align}
which implies that inside the tetrahedron in Fig. \ref{fig1}, there exists an octahedron representing
the region of separable states defined by $|c_1|+|c_2|+|c_3|\leq1$. Therefore, the entangled
Bell-diagonal states reside in the four regions connected to each of the Bell states at the edges. Note that the surfaces of constant entanglement are given by the planes that are parallel to the faces of the octahedron. The four regions of entanglement can be distinguished based on the sign of the three real coefficients. Looking at Fig. \ref{fig1}, it is not difficult to see that in the red and orange entangled
regions (which are connected to the Bell states $|\psi^+\rangle$ and $|\psi^-\rangle$ respectively at the extreme points), the coefficients $c_1$ and $c_2$ have the same sign. On the other hand, in the remaining two entangled regions, they have the opposite sign. Recalling that our global dephasing map leaves $c_3(t)$ and $[c_1(t)+c_2(t)]$ unchanged and taking into account Eq. (\ref{ent}), we conclude that entanglement remains forever frozen throughout the dynamics for the initial states residing in red and orange regions. To put it differently, whereas an initial state will have its $c_3$ constant, it will evolve towards $c_1=c_2$ plane as $t\rightarrow \infty$. Thus, the initial states from the red and orange regions follow a path towards the $c_1=c_2$ plane on which entanglement is constant. Besides, the states from the other two regions cross into the separable octohedron in finite time suffering sudden death, except for the case $c_3=1$, where entanglement decays asymptotically, since these states reach the $c_1=c_2$ plane and the separable region simultaneously.

Motivated by the peculiar phenomenon of time-invariant entanglement, we investigate the violation of the CHSH inequality. After all, just as the presence of entanglement, the violation of the CHSH inequality is also a manifestation of non-classicality. To study the non-local correlations, we introduce the matrix $T=T_{ij }=\textmd{Tr}(\rho(\sigma_{i} \otimes \sigma_{j}))$ and the CHSH operator $B_{CHSH}= \vec{a} \cdot \vec{\sigma} \otimes (\vec{b}+\vec{b'}) \cdot \vec{\sigma} +\vec{a'} \cdot \vec{\sigma} \otimes (\vec{b}-\vec{b'}) \cdot \vec{\sigma},$ where $\vec{a}$, $\vec{a'}$ and $\vec{b}$, $\vec{b'}$ denote the unit vectors indicating the measurements on the first and the second qubits, respectively. The CHSH inequality is then given by $|\textmd{Tr}(\rho B_{CHSH})| \leq 2$. Optimizing over all measurements, the maximum mean value of the CHSH operator reads $\max|\textmd{Tr}(\rho B_{CHSH})|=2\sqrt{M(\rho)}$ \cite{horod}. Here, $M(\rho)=\max_{i<j}\{u_i+u_j\}\leq2$ with $u_j$ being the three eigenvalues of the matrix $U=T^T T$. There exists a choice of measurement setting violating the CHSH inequality if and only if $M(\rho)>1$. For the Bell diagonal states, we have
\begin{equation} \label{chsh}
B(\rho)=2 \sqrt{\max\{c_1^2+c_2^2,c_2^2+c_3^2,c_1^2+c_3^2\}}.
\end{equation}

\begin{figure}[b]
\includegraphics[width=0.48\textwidth]{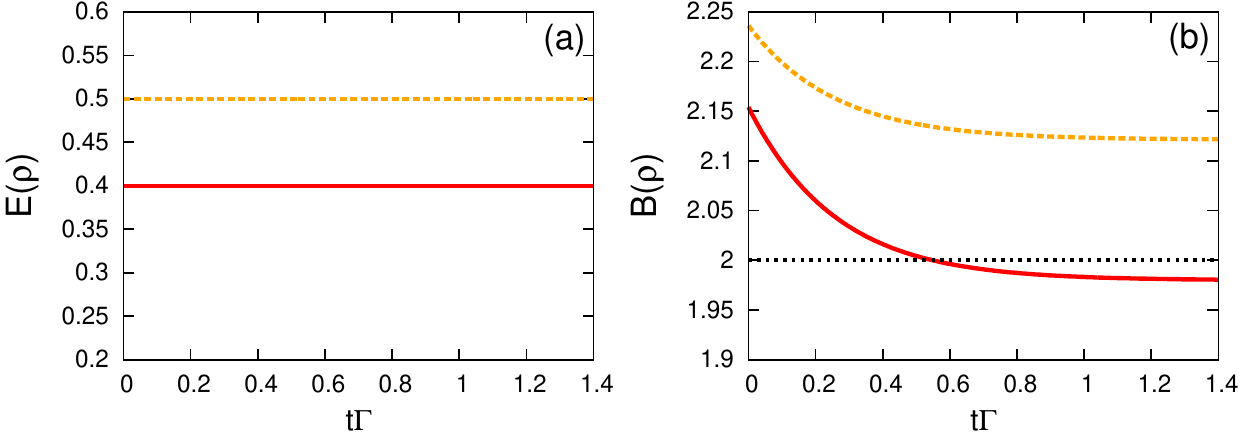}
\caption{Dynamics of (a) entanglement $E(\rho)$ and (b) CHSH violation $B(\rho)$ as a function of the dimensionless time $t\Gamma$. The initial Bell diagonal states are chosen as $c_1=1,c2=0.4,c_3=-0.4$ (red solid line) and $c_1=c_3=-0.5,c2=-1$ (orange dashed line). The black dotted line shows the non-locality threshold.}
\label{fig2}
\end{figure}

In Fig. \ref{fig2}, we display the outcomes of our analysis for entanglement and non-local correlations
with the help of the CHSH inequality considering two different initial Bell-diagonal states under
global dephasing noise. Whereas the red solid line corresponds to the initial state $c_1=1, c_2=0.4,
c_3=-0.4$, belonging to the red entangled region connected to the Bell state $|\psi^+\rangle$ as shown in Fig. \ref{fig1}, the orange dashed line stands for the initial state $c_1=c_3=-0.5, c_2=-1$, which resides inside the orange entangled region connected to the Bell state $|\psi^-\rangle$. We see that entanglement remains time-invariant for both initial states. However, non-local correlations turn out to be susceptible to the effects of the dephasing noise, unlike entanglement. In particular, for the former initial state, non-local correlations vanish in finite time suffering sudden death. Hence, despite the invariance of entanglement, the system might completely lose its non-classicality in terms of violation of the CHSH inequality.

In our experiment, the principal system is represented by a pair of photons, generated via spontaneous parametric downconversion, that are entangled in their polarization degrees of freedom. After a state preparation procedure, the photons are set to move along different arms and pass through quartz plates of adjustable thickness. As each of the photons passes through the quartz plates, its polarization locally couples to its frequency, which acts as the local environment. Such an interaction induces a local unitary transformation $U_i(t_i) |\lambda \rangle \otimes |\omega_i\rangle=e^{in_{\lambda}\omega_{i} t_i} |\lambda \rangle \otimes |\omega_i\rangle$, where $|\lambda \rangle \otimes |\omega_i\rangle$ is the state of the photon in arm $i$ ($i=1,2$) with polarization $\lambda=H,V$ (horizontal or vertical) and frequency $\omega_i$. Here, $t_i$ is the time of interaction and $n_\lambda$ is the refraction index of photons having polarization $\lambda$.

We suppose that the total state of the system and environment can be initially written as a
product state $ |\Psi(0)\rangle=\rho(0) \otimes \int  d\omega_1 d\omega_2 g(\omega_1,\omega_2)
|\omega_1,\omega_2\rangle,$ where $g(\omega_1,\omega_2)$ is the probability amplitude for the photon travelling along arm 1(2) to have frequency $\omega_{1 (2)}$ . The corresponding joint probability distribution reads $P(\omega_1,\omega_2)=|g(\omega_1,\omega_2)|^2$. Then, dynamics of the polarization
state of the photons are given by \cite{Laine2012,Liu2013}
\begin{eqnarray}
\rho(t)=
\left(\begin{array}{cccc}
\rho_{11} & \rho_{12} \kappa_2 & \rho_{13}\kappa_1 & \rho_{14}\kappa_{12} \\
\rho_{21}\kappa_2^* & \rho_{22} & \rho_{23}\Lambda_{12} & \rho_{24}\kappa_1 \\
\rho_{31}\kappa^*_{1} & \rho_{32}\Lambda^*_{12} & \rho_{33} & \rho_{34}\kappa_2 \\
\rho_{41}\kappa^*_{12} & \rho_{42}\kappa^*_1 & \rho_{43}\kappa^*_2 & \rho_{44} \\
\end{array}\right),
\end{eqnarray}
where
$\kappa_1(t)=G(t_1,0)$, $\kappa_2(t)=G(0,t_2)$, $\kappa_{12}(t)=G(t_1,t_2)$ and
$\Lambda_{12}(t)=G(t_1,-t_2)$ with $G(t_1,t_2)=\int  d\omega_1 \omega_2 P(\omega_1,\omega_2)
e^{-i\Delta n (\omega_1 t_1 +\omega_2 t_2 )}$ being the Fourier transform of the joint
probability distribution and $\Delta n = n_V - n_H $ denotes the birefringence.

Assuming that the joint probability distribution of the considered two photons is in a Gaussian form, i.e., $P(\omega_1,\omega_2)=(1/2 \pi \sqrt{\det C}) e^{-\frac{1}{2}(\vec{\omega}-\langle \vec{\omega}
\rangle)^TC^{-1}(\vec{\omega}-\langle \vec{\omega}\rangle)},$
where $\vec{\omega}=(\omega_1,\omega_2)^T$, $\langle \vec{\omega} \rangle=(\langle \omega_1 \rangle
,\langle \omega_2 \rangle)^T$, and $C=C_{ij}=\langle\omega_i\omega_j\rangle-\langle\omega_i\rangle
\langle\omega_j\rangle$ with $\langle\omega_1\rangle=\langle\omega_2\rangle=\omega_0 / 2$ and
$C_{11}=C_{22}=\langle\omega^2_i\rangle- \langle\omega_i\rangle^2$, the decoherence function reads
$G(t_1,t_2)= e^{\frac{i\omega_0}{2}\Delta n (t_1+t_2)-\frac{C_{11}}{2}\Delta n^2(t_1^2+t_2^2+2Kt_1t_2)}$,
with $K=C_{12}/C_{11}$ being the correlation coefficient between the two frequencies. Note that, even though the interactions of the photons with their individual environments are local, the resulting dynamics can be non-local due to the initial correlations between the environments. Let us now set the interactions times to be identical, that is $t_1=t_2=t$, and consider the case of full anticorrelations, corresponding to $K=-1$. We observe that in this case $\Lambda_{12}(t)= e^{-2C_{11}\Delta n^2 t^2}$ and $\kappa_{12}(t)= e^{i\omega_0\Delta n t}$, and thus $|\kappa_{12}(t)|=|\kappa^*_{12}(t)|=1$ at all times, which in turn effectively simulates a global dephasing dynamics creating a decoherence free subspace.

\begin{figure}[t]
\includegraphics[width=0.4\textwidth]{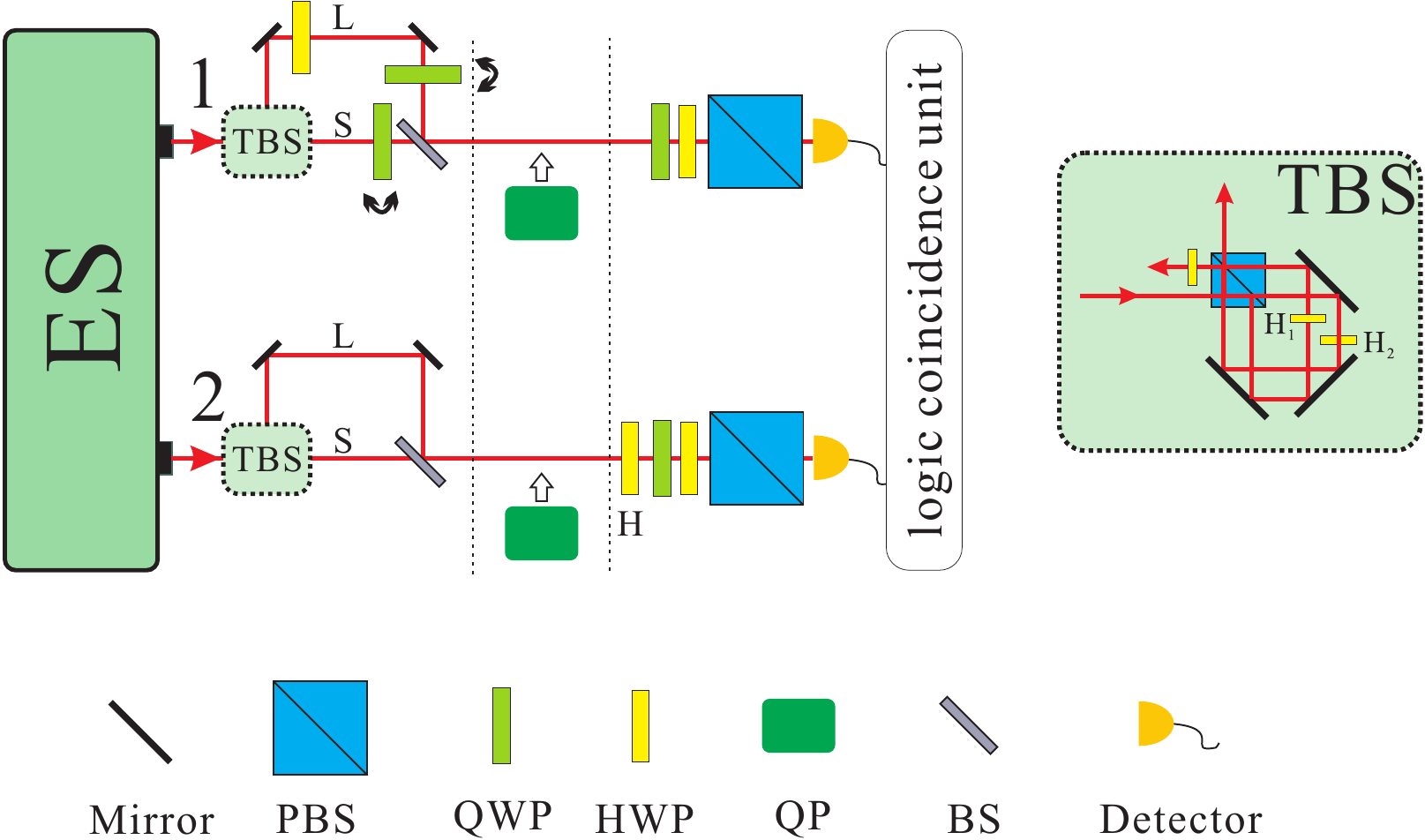}
\caption{Our experimental setup. PBS - polarizing beam splitter, QWP - quarter wave plate, HWP - half wave plate, QP - quartz plate, TBS - tunable beam splitter and ES - entangled photon source. The experiment is performed in three stages that are separated in the figure with two dotted vertical lines. The first stage is the preparation of the Bell-diagonal states. As the second stage deals with the introduction of the global pure dephasing dynamics, the last stage consists of the CHSH measurements and tomography.}
\label{fig3}
\end{figure}

In Fig. \ref{fig3}, we describe our experimental setup. ES is a standard two photon entanglement source
generating the state $|\phi^+\rangle=(|HH\rangle+| VV\rangle)/\sqrt{2}$. After the creation of the
polarization entangled photon pair, each photon is separated by a special designed tunable beam splitter
(TBS) as shown in the dashed inset and travels along 5 m single mode fiber (L) or 1 m single mode fiber (S)
and then combined together at another beam splitter (BS). (The inset TBS contains a polarizing beam splitter
(PBS) and three half wave plates (HWPs). The transmission reflection ratio is well adjusted by the two HWPs
and the relative amplitude of L arm and S arm are well set.) The L part of the photon in arm 1 passes
through a half wave plate set at $45^o$ so that the two photons can be prepared in the Bell diagonal states
$\alpha|\phi^{\pm}\rangle\langle\phi^{\pm}|+(1-\alpha)|\psi^{\pm}\rangle\langle\psi^{\pm}|$. Here, the phase 0 or $\pi$ can be set by tuning the quarter wave plate in arms L and S. In the next stage, each of the photons passes through quartz plates in different arms, and then the photon in arm 2 is rotated by another half wave plate H set at $45^o$, causing the dynamics to resemble the one described in Eq. (\ref{dyn}), by transforming $|\phi^\pm\rangle$ and $|\psi^\pm\rangle$ into each other. Explicitly, after relabelling the elements of the initial state, the dynamics reads
\begin{eqnarray}
\rho(t)=
\left(\begin{array}{cccc}
\rho_{11} & \rho_{12} \kappa_2^* & \rho_{13}\kappa_1 & \rho_{14}\Lambda_{12} \\
\rho_{21}\kappa_2 & \rho_{22} & \rho_{23}\kappa_{12} & \rho_{24}\kappa_1 \\
\rho_{31}\kappa^*_{1} & \rho_{32}\kappa^*_{12} & \rho_{33} & \rho_{34}\kappa^*_2 \\
\rho_{41}\Lambda^*_{12} & \rho_{42}\kappa^*_1 & \rho_{43}\kappa_2 & \rho_{44} \\
\end{array}\right).
\end{eqnarray}
Finally, the resulting two photon state is analysed by the quantum state tomography or CHSH measurements.

In the experiment, we use a CW laser (Toptica, wavelength is 404 nm and power is about 100 mW) to pump two type-I cut 0.3 mm thick BBO to prepare the maximally entangled states. The initial coincidence is about 6000/s and the final coincidence is about 350/s due to the fiber coupling loss and the efficiency of BS. The integration time is 150 s and the total coincidence is about 50000, which gives an error (only calculating the photon number statistic error) about 0.006 and 0.008 for concurrence and CHSH violation, respectively, for the Bell diagonal state with $c_1=1, c_2=0.4, c_3=-0.4$. On the other hand, the integration time is 300 s and the total coincidence is about 100000, giving an error about 0.004 and 0.006 for concurrence and CHSH inequality, respectively, for the Bell diagonal state having $c_1=c_3=-0.5,c2=-1$.

\begin{figure}[t]
\includegraphics[width=0.48\textwidth]{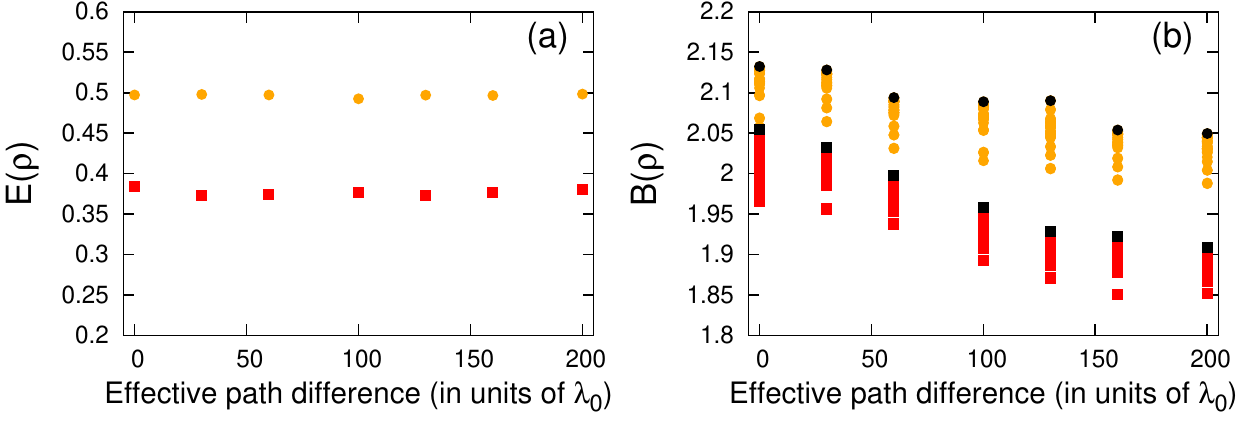}
\caption{(a) Entanglement and (b) CHSH inequality violation versus effective path difference.
The experimentally prepared initial Bell diagonal states are  $c_1=1,c2=0.4,c_3=-0.4$ (red squares) and
$c_1=c_3=-0.5,c2=-1$ (orange circles). In (b), we show the violation of the CHSH inequality for many different measurement settings. The optimal violation is given by the highest points in the y-axis for each state, which are marked with black data points.}
\label{fig4}
\end{figure}

Fig. \ref{fig4} presents the results of our experimental investigation. Specifically, we display entanglement in (a) for states having $c_1=1, c_2=0.4, c_3=-0.4$ and $c_1=c_3=-0.5,c2=-1$ with red squares and orange circles respectively, while (b) shows the degree of CHSH violation for the same pair of states. As we show the CHSH violation for many different measurement configurations in (b), the optimal violation is given by the highest point in the y-axis for each value of the effective path difference. Looking at Figs. \ref{fig2} and \ref{fig4}, we observe that theoretical and experimental results are in good agreement. The comparison between experimental data and theoretical predictions shows a less accurate match for the Bell inequality violation than for the entanglement dynamics. We believe that the reason lies in the fact that the CHSH violation with sensitive correlation measurements is more vulnerable to experimental inaccuracies than quantification of entanglement with state tomography measurements. The experimental data do show, however, conclusive evidence of the fact that, under global dephasing noise, entanglement becomes forever frozen while non-local correlations can suffer sudden death.

\begin{figure}[t]
\includegraphics[width=0.33\textwidth]{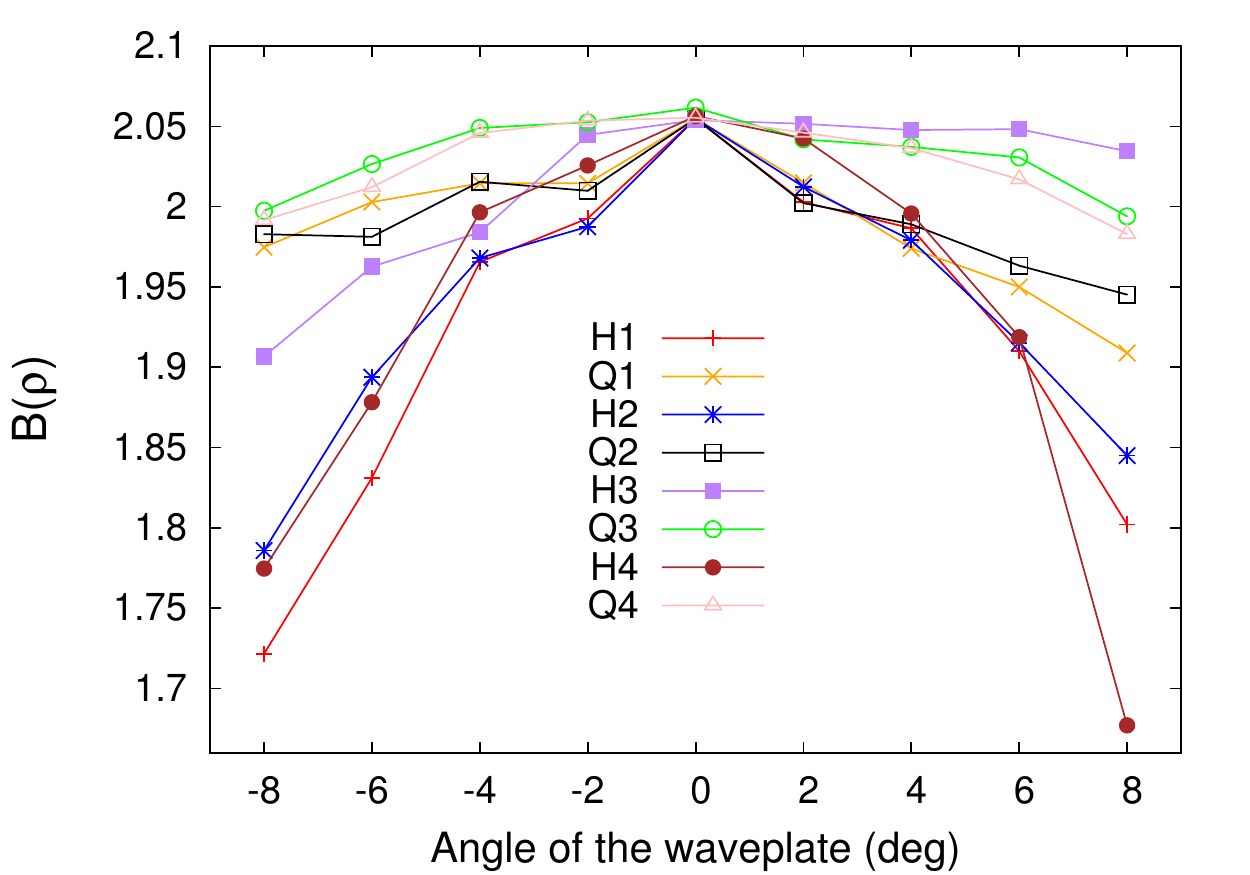}
\caption{Degree of CHSH inequality violation for different measurement configurations around the numerically calculated optimal angles for the Bell diagonal state with $c_1=1, c_2=0.4, c_3=-0.4$. While H1 (H2, H3, H4) form the half wave plate of basis $\vec{a}$ ($\vec{a}'$, $\vec{b}$, $\vec{b}'$), Q1 (Q2, Q3, Q4) form the quarter wave plate of basis $\vec{a}$ ($\vec{a}'$, $\vec{b}$, $\vec{b}'$).}
\label{fig5}
\end{figure}

We lastly elaborate on how we experimentally test the violation of the Bell inequalities. We first reconstruct the density matrix using state tomography. Then, we numerically maximize the degree of violation of the CHSH inequality. Next, we fix all the wave plate angles (corresponding to different measurement bases) at the optimal values and individually change them to check for the maximal experimental violation of the CHSH inequality. We repeat the same procedure for all other states at each value of the effective path difference. In other words, at each time point, we search, within a certain angle interval, for the optimal experimental angles showing maximum violation of the CHSH inequality, in order to compensate for the uncertainties that might occur in the tomography. As an example, we show the results of our analysis for the Bell diagonal state with the coefficients $c_1=1, c_2=0.4, c_3=-0.4$ in Fig. \ref{fig5}. In this case, the optimal violation is indeed given by the angles obtained from the numerical optimization.

In summary, we presented a detailed examination of the dynamics of entanglement and non-local correlations, quantified via concurrence and CHSH inequality violation respectively, under global dephasing for Bell diagonal states. Our results demonstrate that, remarkably, while entanglement can become time-invariant throughout the dynamics for a certain subset of Bell diagonal states, non-local correlations in these states might vanish in a finite time suffering sudden death. Non-local correlations do not seem, according to our investigation, to display a time-invariant behaviour. However, they can reach a non-zero stationary value larger than two which will be maintained during the dynamics, as shown in Figs. \ref{fig4} (b) and \ref{fig2} (b). The existence of time-invariant entanglement and non-locality trapping may pave the way to new strategies, based on reservoir engineering techniques, aimed at exploiting rather than fighting decoherence.

\begin{acknowledgments}
This work was supported by the National Natural Science Foundation of China (Nos. 61327901, 11374288, 11474268, 61225025, 11274289, 11325419), the Strategic Priority Research Program (B) of the Chinese Academy of Sciences (Grant No. XDB01030300), the Fundamental Research Funds for the Central Universities, China (Grant Nos. WK2470000018 and WK2470000022), the National Youth Top Talent Support Program of National High-level Personnel of Special Support Program (No. BB2470000005). GK is supported by the S\~{a}o Paulo Research Foundation (FAPESP) under the grant numbers 2012/18558-5 and 2014/20941-7, and FFF under the grant number 2015/05581-7. FFF is also supported by the National Counsel of Technological and Scientific Development (CNPq) under the grant number 474592/2013-8 and by the National Institute for Science and Technology of Quantum Information (INCT-IQ) under the process number 2008/57856-6. CFL, JP and SM acknowledge financial support by the EU Collaborative project QuProCS (Grant Agreement 641277). JP and SM also acknowledge financial support from the Magnus Ehrnrooth Foundation.
\end{acknowledgments}


\begin{thebibliography}{}
\bibitem{entrev} R. Horodecki, P. Horodecki, M. Horodecki, and K. Horodecki, Rev. Mod. Phys. \textbf{81}, 865 (2009).
\bibitem{discrev}  K. Modi, A. Brodutch, H. Cable, T. Paterek, and V. Vedral, Rev. Mod. Phys. \textbf{84}, 1655 (2012).
\bibitem{openbook} H. -P. Breuer and F. Petruccione, \textit{The Theory of Open Quantum Systems} (Oxford University Press, Oxford, 2007).
\bibitem{entnonmark} B. Bellomo, R. Lo Franco, and G. Compagno, Phys. Rev. Lett. \textbf{99}, 160502 (2007).
\bibitem{entdd} L. Viola and S. Lloyd, Phys. Rev. A \textbf{58}, 2733 (1998); L. Viola, E. Knill, and S. Lloyd, Phys. Rev. Lett. \textbf{82}, 2417 (1999).
\bibitem{entzeno} S. Maniscalco, F. Francica, R. L. Zaffino, N. Lo Gullo, and F. Plastina, Phys. Rev. Lett. \textbf{100}, 090503 (2008).
\bibitem{frozendisc} P. Haikka, T. H. Johnson, and S. Maniscalco, Phys. Rev. A {\bf 87}, 010103(R) (2013).
\bibitem{frozenent1} G. Karpat and Z. Gedik, Phys. Lett. A \textbf{375}, 4166 (2011).
\bibitem{frozenent2} E. G. Carnio, A. Buchleitner, and M. Gessner, Phys. Rev. Lett. \textbf{115}, 010404 (2015).
\bibitem{bell} J. S. Bell, Physics \textbf{1}, 195 (1964).
\bibitem{gisin} N. Gisin, Phys. Lett. A \textbf{154}, 201 (1991).
\bibitem{werner} R. F. Werner, Phys. Rev. A \textbf{40}, 4277 (1989).
\bibitem{chsh} J. F. Clauser, M. A. Horne, A. Shimony, and R. A. Holt, Phys. Rev. Lett. \textbf{23}, 880 (1969).
\bibitem{kofman} A. G. Kofman and A. N. Korotkov, Phys. Rev. A \textbf{77}, 052329 (2008).
\bibitem{entnonloc} B. Bellomo, R. Lo Franco, and G. Compagno, Phys. Rev. A \textbf{78}, 062309 (2008).
\bibitem{entchsh} L. Mazzola, B. Bellomo, R. Lo Franco, and G. Compagno, Phys. Rev. A \textbf{81}, 052116 (2010).
\bibitem{entchsh2} B. Bellomo, R. L. Franco and G. Compagno, Adv. Science Lett. \textbf{2}, 459 (2009).
\bibitem{entchsh3} C. Bengtson, M. Stenrup, and E. Sj\"{o}qvist, Int. J. Quantum Chem. \textbf{116}, 1763 (2016)
\bibitem{Shu2010} J.-S. Xu {\textit {et al.}}, Phys. Rev. Lett. \textbf{104}, 100502 (2010).
\bibitem{key} A. Acin, N. Gisin, and L. Masanes, Phys. Rev. Lett. \textbf{97}, 120405 (2006)
\bibitem{yueb} T. Yu, J.H. Eberly, Phys. Rev. B \textbf{68}, 165322 (2003).
\bibitem{conc} W. K. Wootters, Phys. Rev. Lett. \textbf{80}, 2245 (1998).
\bibitem{horod} M. Horodecki, P. Horodecki, and R. Horodecki, Phys. Lett. A \textbf{200}, 340 (1995).
\bibitem{Laine2012} E.-M. Laine {\textit {et al.}} Phys. Rev. Lett. \textbf{108}, 210402 (2012),
Erratum: {\textit{ibid.}} 111, 229901 (2013).
\bibitem{Liu2013} B.-H. Liu {\textit{et al.}}, Sci. Rep. \textbf{3}, 1781 (2013).

\end{thebibliography}
\end{document}